\documentclass[conference,a4paper]{APSIPA2021}
\usepackage{cite,balance}
\usepackage{multirow, multicol,graphicx}
\usepackage{amsmath}
\usepackage[psamsfonts]{amssymb}
\usepackage{amsxtra}
\usepackage{threeparttable}
\begin{document}

\title{Significance of Data Augmentation for Improving\\ Cleft Lip and Palate Speech Recognition}

\author{%
\authorblockN{%
Protima Nomo Sudro\authorrefmark{1}, Rohan Kumar Das\authorrefmark{3}, Rohit Sinha\authorrefmark{1}, S. R. Mahadeva Prasanna\authorrefmark{2}
}

\authorrefmark{1}Indian Institute of Technology Guwahati, Guwahati, India\\

\authorrefmark{2}Indian Institute of Technology Dharwad, Dharwad, India\\

\authorrefmark{3}Fortemedia Singapore, Singapore\\
E-mail:\{protima, rsinha\}@iitg.ac.in, prasanna@iitdh.ac.in, rohankd@ieee.org}

  


\maketitle
\thispagestyle{empty}
\begin{abstract}

The automatic recognition of pathological speech, particularly from children with any articulatory impairment, is a challenging task due to various reasons. The lack of available domain specific data is one such obstacle that hinders its usage for different speech-based applications targeting pathological speakers. In line with the challenge, in this work, we investigate a few data augmentation techniques to simulate training data for improving the children speech recognition considering the case of cleft lip and palate (CLP) speech. The augmentation techniques explored in this study, include vocal tract length perturbation (VTLP), reverberation, speaking rate, pitch modification, and speech feature modification using cycle consistent adversarial networks (CycleGAN). Our study finds that the data augmentation methods significantly improve the CLP speech recognition performance, which is more evident when we used feature modification using CycleGAN, VTLP and reverberation based methods. More specifically, the results from this study show that our systems produce an improved phone error rate compared to the systems without data augmentation.

\end{abstract}

\vspace{2mm}
\noindent\textbf{Index Terms}: CLP speech, data augmentation, automatic speech recognition 

\section{Introduction}

Disordered speech is collectively used to refer to the speech that deviates both in intelligibility and quality~\cite{kummer2013cleft, peterson2001cleft}. The deviations may be caused by structural and functional deformation of the articulatory system~\cite{grunwell2001speech}. Such speech deviations affect the speech understandability by unfamiliar listeners due to a communication gap between impaired and normal speakers~\cite{whitehill2002assessing}. This kind of mismatch in the acoustic characteristics between the normal and impaired speech makes the automatic recognition of the disordered speech a challenging task. Besides the mismatch in acoustic characteristics, the type of speech distortions such as hypernasality, articulation error and levels of severity of the speech distortions among the pathological speakers impose additional challenges in case of disordered speech recognition~\cite{maier2006intelligibility,vucovich2017automated,yilmaz2018articulatory, shor2019personalizing}. 

Various studies~\cite{yu2017recent, cui2015data, kanda2013elastic} reported in automatic speech recognition (ASR) literature highlight that a model built using a large amount of speech data often achieves high recognition accuracy and robust performance. However, when it comes to pathological speech, collecting a large amount of data is very challenging as the speakers face difficulty in speaking for a long time, apart from having sufficient numbers of speakers for a particular case of pathological speech. In addition, the time alignment of the collected data is painstaking. On account of data scarcity and acoustic variation, the disordered speech recognition systems are noted to yield relatively inferior recognition performances when compared with the normal speech recognition systems. 



In the context of normal ASR systems, the data augmentation approaches are found to be effective for dealing with data scarcity as well as increasing robustness under adverse noisy conditions~\cite{park2019specaugment}.
The ASR systems play an important role for various speech based applications in the recent years, particularly with smart-home devices that are operated by voice commands and dialogues~\cite{bajpai2019smart, moller2006evaluating}.
However, as most of those speech based applications are trained with speech data from normal speakers, the practical usability of such systems for pathological speakers is limited. To deal with the data scarcity and domain mismatch issues, data augmentation approaches have been exploited for improving the performance of disordered speech recognition~\cite{geng2020investigation,jiao2018simulating}. Among them, the dysarthric speech has gained relatively more attention in the recent years. We find the explorations on cleft lip and palate (CLP) speech in the context of improving ASR is limited, which motivated us to focus in this study. 


The CLP is a congenital disorder affecting the craniofacial region including the articulatory system~\cite{kummer2013cleft,grunwell2001speech}. Due to the impaired articulatory system, different speech disorders are noted, which are broadly categorized into hypernasality, hyponasality, articulation error (misarticulation), and voice disorder~\cite{kummer2013cleft, peterson2001cleft}. The speech disorders are caused by velopharyngeal dysfunction, oro-nasal fistula and mislearning~\cite{moore1975phonetic}. They impact the intelligibility and quality of CLP speech based on the degree of the speech disorder severity~\cite{kummer2013cleft,scipioni2009intelligibility,maier2006intelligibility}. 

The impairment in the articulatory system affects the acoustic characteristics of the speech. The individuals with CLP exhibit deviant burst evidence, formant transitions, and spectral attributes compared to the normal speech. The acoustic characteristics of the speech sounds are also distorted because of the unintentional production of nasalized vowels and nasal cognates, as certain voiced stops share the same place of articulation with nasal consonants~\cite{henningsson2008universal}. When the distorted speech is input to the speech system, the performance accuracy decreases significantly. The speech systems trained using normal speech typically ignore the unnatural variations of pathological speech~\cite{young2010difficulties}. However, such features are paramount in pathological speech and are not encoded in the statistical frameworks of the speech systems~\cite{rudzicz2013adjusting}.

The distortions in CLP speech can be corrected using clinical interventions, namely, surgery, prosthetics, and speech therapy. Despite the advantage of the interdisciplinary team in the clinical settings and adverse circumstances of the speakers with CLP, many speakers still produce deviant speech compared to the normal speech. For the speakers whose speech distortions cannot be improved, may face difficulty in using various speech based applications effectively as desired. Hence, besides clinical intervention, different approaches from the perspective of signal processing research can be explored for improving the usability of speech based applications. Accordingly, the study devotes to generate variations of CLP speech samples using various data augmentation techniques and then to identify the methods having a higher impact on improving performance of ASR systems.


The rest of the paper is organized as follows. Section~\ref{literature} discusses the literature related to data augmentation studies carried out for improving the performances of normal and disordered speech recognition. In Section~\ref{propose}, we explain the details of various data augmentation methods explored this work. Section~\ref{expt} presents the experimental setup of the studies. The observations and results of conducted studies are reported in Section~\ref{results}. Finally, Section~\ref{sumry} concludes this work.

\section{Related Work}
\label{literature}

Data augmentation refers to the process of generating variations of input data to increase the number of examples for a given dataset. It has been proved effective for both in case of speech and image processing domains~\cite{cirecsan2011high,krizhevsky2012imagenet,lecun1998gradient}. Several studies reported the use of data augmentation approaches to deal with data scarcity in low resource speech based tasks. It is also exploited for useful  applications of speech processing systems such as ASR, transcription of multi genre media archives, children speech recognition, anti-spoofing and modeling large scale complex models~\cite{bell2012transcription,shahnawazuddin2020voice,jaitly2013vocal,ICASSP2021_rkd,das21_asvspoof}. Literature shows a wide range of data augmentation techniques that are explored in the context of speech processing, namely, vocal tract length perturbation, spectral distortion, voice conversion based on adversarial training, speaking rate modification, pitch modification, stochastic feature mapping, and spectrogram deformations, text-to-speech data augmentation, pseudo-label augmentation~\cite{jaitly2013vocal,shahnawazuddin2020voice,jiao2018simulating,kanda2013elastic,cui2015data,park2019specaugment,tsunoo21_interspeech}.

Motivated by the capability of data augmentation methods in dealing with data scarcity and attaining improved accuracy, researchers have explored the same for disordered speech recognition system~\cite{geng2020investigation,jiao2018simulating}. Most of the studies attempted dysarthric speech recognition system, where the variations of the distorted speech are generated using some of the augmentation techniques stated above. In those studies, speech characteristics of normal speakers are transformed to that of disordered speakers. In addition, the augmentation techniques were exploited based on the disordered speech phenomena such as moderate speaking rate of the normal speakers are matched to the slow rate of the disordered speakers. In some cases, normal speakers' pitch is linearly transformed to the reduced pitch variation of the disordered speakers. Similarly, other characteristics of disordered speakers are also considered while transforming normal speech into disordered speech~\cite{xiong2019phonetic}. All these studies suggest that the data augmentation approaches result in improved recognition performance for disordered speech.


Along similar direction, this study intends to exploit data augmentation approaches and transform the characteristics of normal children's speech based on the deviated acoustic characteristics of CLP speech. We note that prominent distortions like nasalization and articulation errors are found in CLP speech. Therefore, the spectral deviations caused by nasalization or articulation error can be mapped into the normal speakers' spectral characteristics to generate variations of CLP speech using normal speech data. A few known data augmentation approaches reported in the literature are also investigated in this study to quantify their relative impacts on the recognition performances.

\section{Data Augmentation for CLP Speech}
\label{propose}

The study aims to improve CLP speech recognition for serving the need for using various speech-based applications. The following subsections describe the data augmentation techniques explored in this study followed by their observations.


\subsection{Vocal tract length perturbation}
The vocal tract length perturbation (VTLP) method is one among the commonly used data augmentation approaches. The speakers with CLP exhibit disordered speech by either changing the place of articulation (PoA) or manner of articulation (MoA) or both. The speakers change the PoA in response to impaired structure in their articulatory system. As a result the vocal tract parameters are distorted. Hence, in this work, the VTLP approach is used as one of the augmentation approaches for investigating its affect in the case of CLP speech recognition. 

In this method, a random warp factor ($\alpha$) is used to add variations in the input data. For each utterance in the training set, a random $\alpha$ is generated and then the original frequency $f$ is warped into a new frequency $f'$ using the technique reported in~\cite{lee1998frequency} where, 
\begin{equation}
X'(f)= X(\alpha f)
\end{equation}
The $\alpha$ values are chosen from a discrete set of values ranging between $[0.9 - 1.1]$. The VTLP approach modifies the spectral characteristics of the input speech signal while preserving the fundamental frequency and duration of the signal. 

\subsection{CycleGAN based speech feature modification}

Another way of data augmentation is accomplished by transforming the acoustic characteristics of normal speech towards the CLP speech characteristics and vice versa. The articulatory impairment distorts the acoustic characteristics of CLP speech. Therefore, mapping the CLP speech speech into that of normal speech will generate a different variation of CLP speech which will resemble the normal speech characteristics. Hence, we explore the cycle-consistent adversarial network (CycleGAN) adversarial training for transforming the acoustic characteristics as described in\footnote{https://github.com/leimao/Voice-Converter-CycleGAN}~\cite{kaneko2018cyclegan}. The CycleGAN is widely used for non-parallel voice conversion (VC) and it has shown its effectiveness for various other applications~\cite{kaneko2018cyclegan,yeh2018rhythm,fang2018high,sudro2021enhancing}. A CycleGAN method is based on a combination of two generators $G$ and $F$ and two discriminators $D_X$ and $D_Y$, respectively. The generator $G$ is a function that transforms the distribution $X$ into distribution $Y$, whereas the generator $F$ transforms the distribution $Y$ into distribution $X$. On the other hand, the discriminator $D_X$ distinguishes between the distribution of $X$ and distribution of $\hat{X}=F(Y)$. In contrast, the discriminator $D_Y$ distinguishes $Y$ from $\hat{Y}=G(X)$. The CycleGAN model learns the mapping function from the training samples, which comprises of source $\{x_i\}^{N}_{i=1} \in X$ and target $\{y_i\}^{N}_{i=1} \in Y$ samples. In CycleGAN approach, both the models are trained simultaneously using the objective function that consists of two losses: adversarial loss and cycle-consistency loss. With adversarial training of the generator and discriminator models, it is expected that  the generated normal speech samples become indistinguishable from CLP speech and vice versa. 

\subsection{Reverberation}

As the present work is motivated towards improving CLP speech recognition for its effective use of speech based applications, there are possible scenarios such as room reverberations and acoustics that may impact the intelligibility and quality of CLP speech. Therefore, we consider room reverberation as one of the strategies for data augmentation. We generate the 
reverberant signal by convolving the input source signal with the room's acoustic impulse response by using Roomsimove\footnote{http://homepages.loria.fr/evincent/software/Roomsimove\_1.4.zip}~\cite{vincent2008roomsimove}.


\subsection{Pitch modification}

The pitch or fundamental frequency, $F_0$ is one of the important acoustic cues of speech signal. Along with various speech disorders, voice disorder is also observed in some of the individuals with CLP~\cite{kummer2013cleft}. In response to velopharyngeal dysfunction, hyperadduction can occur resulting in high pitch. Therefore, the pitch variation is also considered as another augmentation technique for this study. The pitch is modified by using a factor $\beta$, which is given by, 
\begin{equation}
\hat{F_0}=F_0 \times \beta
\end{equation}
where $\beta = \frac{\sigma{F_{0_{CLP}}}}{\sigma{F_{0_{N}}}}$, it is defined as the ratio of standard deviations between the average pitch of CLP speakers and the normal speakers. The pitch modification is performed by using open-source Praat Vocal Toolkit\footnote{https://www.fon.hum.uva.nl/praat/}~\cite{corretge2012praat}, which uses pitch-synchronous overlap and add (PSOLA) method to synthesize the transformed signal~\cite{moulines1990pitch}. It is noted that the duration of the original signal is preserved, while performing pitch modification. 
 
\subsection{Speaking rate transformation}

In CLP speech due to articulatory impairment, speech rate variation is also observed~\cite{jones1985effect}. Hence, we explore the impact of speaking rate transformation in CLP speech recognition. Here, each of the normal speech samples is matched to that of the CLP speech in terms of the speaking rate. The normal speech rate is mapped to the CLP speech rate using dynamic time warping~\cite{myers1980performance}. The time-aligned signals are then synthesized using overlap and add (OLA) method. Speaking rate transformation leads to the change in duration of the signal, while preserving the spectral characteristics and pitch of the normal speaker.

\subsection{Observations from data augmentation on CLP speech}

We are now interested to study the impact of various data augmentation methods explored in this study. For this analysis, we consider an example of CLP speech and then apply the different data augmentation methods to observe the corresponding speech waveforms and spectrograms. 

\begin{figure*}[t!]
\begin{center}
\includegraphics[width=0.9 \textwidth]{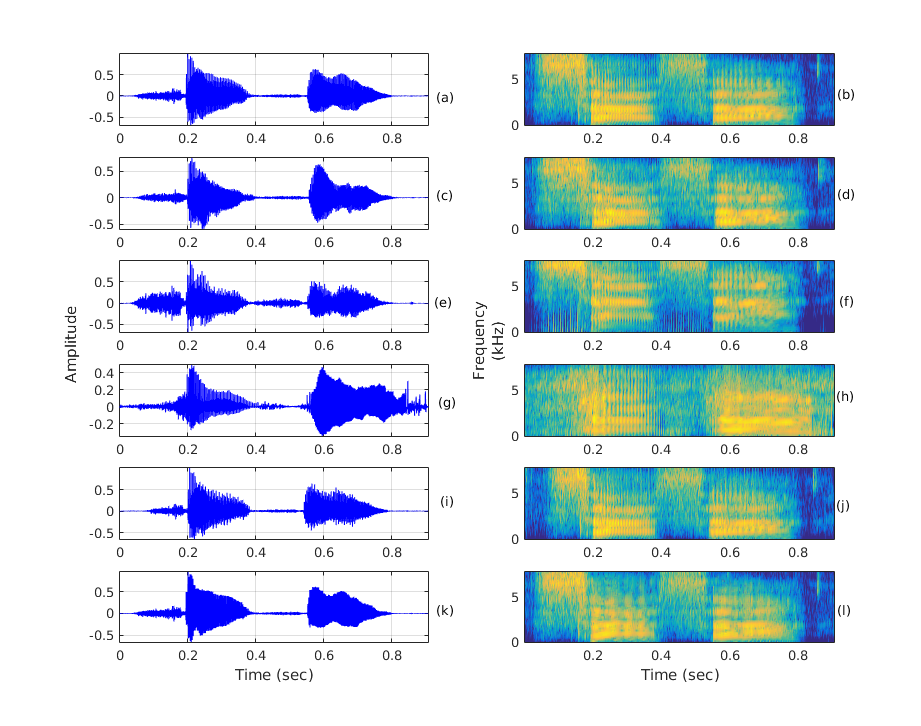} 
\end{center}       
		\vspace{-0.7cm}
		\caption{ Waveforms and spectrograms of various transformed speech for a CVCV word /sasa/. (a)-(b) original normal speech signal (c)-(d) reverberated version, (e)-(f) vocal tract length warped version, (g)-(h) CycleGAN generated version, (i)-(j) speaking rate modified version, and  (k)-(l) pitch modified version.}
		\vspace{-0.4cm}
		\label{s1}
\end{figure*}

Fig.~\ref{s1} shows the comparison of speech waveforms and spectrograms for original speech and that generated using various data augmentation approaches. From Fig.~\ref{s1} (c) and (d), we observe that additional interfering acoustic components are imposed on the original speech signal due to reverberation. With frequency warping factor, $\alpha =0.91$ applied in Fig.~\ref{s1} (e) and (f), it is observed that the high energy spectral components are sampled more sparsely as compared to original normal speech in Fig.~\ref{s1} (a) and (b). Further, with CycleGAN approach shown in Fig.~\ref{s1} (g) and (h), it is observed that the speech characteristics vary significantly because of mapping the normal speech features to that of the CLP speech. The perturbations are observed in the form of indistinguishable formants in the lower frequency region for voiced sounds and low energy concentration of the high frequency signals such as fricative /s/ in the figure.

Considering the speaking rate modification shown in Fig.~\ref{s1} (i) and (j), the duration of the speech signal is reduced to match the corresponding speaking rate of a gender and age-matched CLP speaker. It is noted that other speech parameters are preserved while performing speaking rate modification. Further, from pitch modification of the signal with a modification factor $\beta =1.3$ as shown in Fig.~\ref{s1} (k) and (l), it is observed that the shift in $F_0$ with $65$~Hz preserves the spectral characteristics and the duration of the original signal. Overall, from Fig.~\ref{s1}, it is observed that various data augmentation methods alter the speech parameters such as pitch, duration, intelligibility, quality, and location of high energy spectral contents of the input signal. On the other hand, certain other aspects of speech like speaking style is preserved when the pitch and speaking rate is modified. Similarly, the $F_0$ is kept unaltered while transforming the speech features using CycleGAN based approach.

\subsection{Perceptual evaluation}
The database for this work is acquired from Kannada speaking children. The speech data was collected in a sound-treated room using a Bruel and kj\ae r speech level meter~\cite{slm_recording}. All the speech data were collected in the All India Institute of Speech and Hearing (AIISH), Mysuru, India. The database consists of $72$ speakers, where $39$ speakers ($23$ male and $16$ female) are the individuals with CLP and rest $33$ speakers ($12$ male and $21$ female) are individuals with normal speech. 

The age of CLP and normal participants are $9 \pm  2$ years (mean $\pm$ SD) and $10~\pm~2$ years (mean $\pm$ SD), respectively. The CLP speech characteristics exhibits a combination of speech disorders, namely, hypernasality, articulation errors, and nasal air emission. The transcription and rating of the severity levels of the CLP speech samples are obtained from $3$ expert speech language pathologists (SLP) from AIISH. The SLPs have an experience of more than five years who routinely work with the speakers with CLP. The SLPs were asked to provide rating and transcribe the speech samples based on the instructions given in~\cite{henningsson2008universal}. According to the rating strategy reported in~\cite{henningsson2008universal}, the SLPs provide deviation scores on a scale of $0$ to $3$, where $0=$ close to normal, $1=$ mild deviation, $2=$ moderate deviation, and $3=$ corresponds to severe deviation.

The augmented speech signals are analyzed using subjective evaluation to determine the impact of different data augmentation methods. For this purpose, the perturbed normal and CLP speech samples are mixed and then randomly presented to the listeners. While providing the speech samples for perceptual similarity, the reference speech samples for both the normal and CLP speech are also provided. A total of $15$ participants have performed the perceptual judgement task. For each type of augmentation, $5$ normal reference, $5$ augmented normal, $5$ CLP reference, and $5$ augmented CLP speech samples are  presented for the evaluation. Therefore, a total of $20$ words are provided for each augmentation approach. We note that for CycleGAN method, $5$ normal transformed samples, i.e., from normal to CLP (${\text{N}}_{\text{N} \rightarrow \text{C} }$) and $5$ CLP transformed samples, i.e., from CLP to normal speech ( ${\text{C}}_{\text{C} \rightarrow \text{N}}$) along with their reference signals are presented for evaluation. Overall, 100 speech samples (original reference and augmented) are presented for perceptual evaluation. However, it can be noted that in Table~\ref{perceptual}, the evaluations are shown for augmented speech samples, i.e., 50 words only.



\begin{table}[t!]
	\centering
	\caption{Perceptual similarity (\%) and MOS values for different data augmentation methods calculated over the normal (N) and CLP (C) speech signals. NP stands for no preference.}
	
	\label{perceptual}
	\scalebox{0.88}{
\begin{tabular}{cccccc}
\hline \hline
\multirow{2}{*}{{\bf Method}}&\multirow{2}{*}{{\bf \# Samples}} & \multicolumn{3}{c}{{\bf Similarity (\%)}} & \multirow{2}{*}{{\bf MOS}} \\ \cline{3-5}
                                                 &     & {\bf Normal}      & {\bf CLP}     & {\bf NP}                                   &             \\ \hline 
VTLP                   & 5 N + 5 C                                   & 40          & 32       & 28                                     &  1.85               \\ \hline
CycleGAN               & 5 N {\tiny $_{\text{N} \rightarrow \text{C} }$ }  + 5 C {\tiny $_{\text{C} \rightarrow \text{N} }$ }       & 43    & 37       & 20                      &  2.03             \\\hline

Reverberation          & 5 N + 5 C                                   & 42          & 38       & 20                                     &  2.00          \\ \hline
Speaking rate          & 5 N + 5 C                                   & 46          & 48       & 6                                      &  2.68          \\\hline
Pitch                  & 5 N + 5 C                                   & 47          & 44       & 9                                      &  2.91        \\ \hline

{\bf Total}            & {\bf 25 N + 25 C}                           & {\bf 43.6}  & {\bf 39.8}& {\bf 16.6}                            &  {\bf 2.3}       \\ \hline 
No augmentation        & 25 N                                        & 98          & -        & 2                                      &  4.73        \\  \hline 
No augmentation        & 25 C                                        & 8           & 88        & 4                                      &  2.65        \\  \hline \hline
\end{tabular}
}
 \vspace{-0.5cm}
\end{table}


In the evaluation process, two different tasks were assigned to the listeners. The first task is designed to provide preferences for each augmented speech sample whether they are perceptually similar to the speech of normal speaker or CLP speaker. If the listener found difficult to make any judgement, they can select no preference option. In the perceptual evaluation, the second task of the listener aims to examine whether the augmentation approaches have altered the overall quality of the speech significantly. For this, the mean opinion score (MOS) values are computed using the rating scale ranging from $1$ to $5$ ($1 = $ bad, $2 = $ fair, $3 = $ good, $4 =$ very good, $5 =$ excellent. The perceptual evaluation results are shown in Table~\ref{perceptual}.

\section{Experimental Setup}
\label{expt}

Although the database consists of non-meaningful CVCV words and meaningful short phrases, most of the augmentation experiments were carried out using isolated CVCV words only. The exceptional case is explored in CycleGAN based approach, where the transformation model is collectively trained using the mel-cepstral coefficients extracted from each frame of the short phrases from normal and CLP speakers. Using the trained model, the features of the CVCV words from normal speech are transformed into that of CLP speech features. Similarly, another model is trained for transforming the CLP speech characteristics into normal speech characteristics. For the sake of creating more input variations, the fundamental frequency of the normal speech and CLP speech are preserved while modifying the spectral characteristics of the signal using CycleGAN method. 

Table~\ref{perceptual} shows that approximately $43.6\%$ of the normal augmented speech signals are perceived as normal  and approximately $39.8 \%$ of the CLP augmented signals are perceived as CLP speech. In addition, we notice that a significant number of speech samples are different (i.e., no judgement were made for these samples) from normal and CLP reference. Considering the MOS, an average score of $2.3$ is observed. The lower MOS value is because the perturbations have altered the perceptual quality of the speech signals to a certain extent.

Another factor for overall lower MOS value is that equal amount of normal and CLP speech samples are presented for the perceptual evaluation. Despite the higher MOS value of normal speech, perturbations and lower MOS of CLP speech have significantly reduced the overall MOS value.

\begin{table}[!tbh]
\centering
	\caption{Impact of different data augmentation methods in PER (\%) on the test set over CLP speech signals }
	\label{per_clp}
	\scalebox{0.9}{
\begin{tabular}{cccc}
\hline\hline
Method        & \begin{tabular}[c]{@{}c@{}}Amount of \\ augmented data\end{tabular} & \begin{tabular}[c]{@{}c@{}}\# Training\\ utterances\end{tabular} & PER (\%) \\ \hline
Not applied   & -                                                                   & 5296                                                             & 50.68   \\
VTLP          & 2$ \times $                                              & 10,592                                                           & 42.95   \\
CycleGAN      & 2$ \times $                                              & 10,592                                                           & {\bf 41.09}   \\
Reverberation & 2$ \times $                                              & 10,592                                                           & 42.10   \\
Speaking rate & 2$ \times $                                              & 10,592                                                           & 48.67   \\
Pitch         & 2$ \times $                                              & 10,592                                                           & 43.91   \\ \hline\hline
\end{tabular}
}
\vspace{-2mm}
\end{table}

\begin{table}[!tbh]
\centering
	\caption{Impact of different data augmentation methods in PER (\%) on the test set over normal speech signals }
	\label{per_nrml}
	\scalebox{0.9}{
\begin{tabular}{cccc}
\hline\hline
Method        & \begin{tabular}[c]{@{}c@{}}Amount of \\ augmented data\end{tabular} & \begin{tabular}[c]{@{}c@{}}\# Training\\ utterances\end{tabular} & PER (\%) \\ \hline
Not applied   & -                                                                   & 5297                                                             & 21.59   \\
VTLP          & 2$ \times$                                              & 10,594                                                           & 16.81   \\
CycleGAN      & 2$ \times $                                              & 10,594                                                           & {\bf 14.84}   \\
Reverberation & 2$ \times $                                              & 10,594                                                           & 15.68   \\
Speaking rate & 2$ \times $                                             & 10,594                                                           & 17.46   \\
Pitch         & 2$ \times $                                              & 10,594                                                           & 16.95   \\ \hline\hline
\end{tabular}
}
\vspace{-2mm}
\end{table}

\begin{table}[!tbh]
\centering
	\caption{Impact of different data augmentation methods in PER (\%) on the test set over normal and CLP speech signals }
	\label{per_clp_nrml}
	\scalebox{0.9}{
\begin{tabular}{cccc}
\hline\hline
Method        & \begin{tabular}[c]{@{}c@{}}Amount of \\ augmented data\end{tabular} & \begin{tabular}[c]{@{}c@{}}\# Training\\ utterances\end{tabular} & PER (\%) \\ \hline
Not applied   & -                                                                   & 10,593                                                           & 58.49   \\
VTLP          & 2$\times$                                             & 21,186                                                           & 49.21   \\
CycleGAN      & 2$ \times $                                              & 21,186                                                           & {\bf 44.52}   \\
Reverberation & 2$ \times $                                              & 21,186                                                           & 48.98   \\
Speaking rate & 2$ \times $                                              & 21,186                                                           & 56.02   \\
Pitch         & 2$ \times $                                              & 21,186                                                           & 50.34   \\ \hline\hline
\end{tabular}
}
\vspace{-4mm}
\end{table}

\section{Results and Discussion}
\label{results}



This study focuses on investigating the ASR performance for CLP speech using various augmentation approaches. Hence, the deformed speech samples are augmented with the original speech samples and then analyze the ASR results. In the current work, double of data size is augmented to maintain the acoustic characteristics of both the original and augmented data equally. As the speech data for this study are in the Kannada language, a Kannada ASR system is developed using Kaldi speech recognition toolkit\footnote{https://github.com/kaldi-asr/kaldi}~\cite{povey2011kaldi}. In the experiment, the hybrid deep neural network (DNN) acoustic model was implemented. All the speech samples are downsampled to $16$~kHz before processing. The ASR system performance for various data augmentation methods is measured using phone error rate (PER) metric. The PER is calculated based on the general form of error rate (ER) computation given by,  $ \text{ER =(substitutions+deletions+insertions)/(total phones)}$.

Tables~\ref{per_clp}-\ref{per_clp_nrml} reports the PER for different studies on the test set along with number of training utterances, which is more when various data augmentations are applied. In Table~\ref{per_clp}, the PER values obtained from CLP speech are reported. For training, $37$ CLP speakers data are used and $2$ CLP speakers data are used for testing. Three-fold cross validation is performed and the final measure was obtained by averaging across all folds.  During three-fold cross validation, the test set comprises of female-female, male-male and female-male combinations, respectively.  The test set consists of an average of $279$ utterances in each fold. Table~\ref{per_nrml} reports the PER values obtained from normal speech. In this case, $31$ normal speakers data are used for training and $2$ normal speakers data are used for testing. Here also three-fold cross validation is performed and the combinations are same as stated before for the CLP speech recognition. In Table~\ref{per_clp_nrml}, the PER values are obtained from the combination of normal and CLP speech. A total of $70$ speakers (normal and CLP) data are used for training the system and $2$ CLP speakers data are used for testing. Three-fold cross validation and aforementioned combinations are used to obtain the final PER value.

For comparing the performances of the augmentation approaches, the PER values obtained for the original speech signals forms the baseline measures. From Tables~\ref{per_clp}-\ref{per_clp_nrml}, it is observed that all explored data augmentation approaches have yielded a significant improvement in the PER performances. In Table~\ref{per_clp_nrml}, a significant PER improvement of around $14$\% over the baseline is observed when the speech samples are augmented with CycleGAN based transformed data. This is attributed to the increase in the speech samples that exhibit very close characteristics to CLP speech. From tables~\ref{per_clp}-\ref{per_clp_nrml}, it is also observed that CycleGAN based augmentation yielded respectively lower PER value compared to other augmentation methods. In addition, we find that VTLP and reverberation based data augmentation method also contributes to improve PER significantly. On the other hand, the improvements in PER by pitch and speaking rate modification are relatively less compared to other techniques investigated in this work.

As discussed, this paper reports CLP speech recognition study using CVCV words. In contrast to the non-sensical CVCV words, for a realistic scenario, the recognition performance for meaningful sentences, words, and phrases are very relevant, which we intend to explore in the future. Additionally, in this study, a combination of CLP speech disorders are considered, which may hinder the recognition accuracy. Hence, future studies in this direction focusing on generalizing the variability of the speech disorders in this population or consideration of specific type of the speech disorder and then study its impact in the recognition performances deserves attention. In this work, we study CLP children speech that exhibit highly varying speech characteristics. In this regard, we can explore text-to-speech based data augmentation method to generate various speech styles and observe its impact in the recognition performance. Additionally, we can also investigate pseudo-label based data augmentation method to study the impact of label errors and biases in distorted CLP speech.

\section{Conclusion}
\label{sumry}

This work investigates a few data augmentation approaches for improving the CLP speech recognition. The data augmentation methods include VTLP, reverberation, speaking rate, pitch modification, and speech feature modification using CycleGAN. We study them using CLP and normal speech data collected in Kannada language. The studies show that CycleGAN based speech feature modification method improves the recognition performance significantly compared to other data augmentation techniques considered in this study. In addition, VTLP and reverberation based approaches show relatively better results than pitch and speaking rate modification methods.


\section{Acknowledgement}

The authors would like to thank Dr. M. Pushpavathi and Dr. Ajish Abraham, AIISH Mysore, for providing insights about CLP speech disorder. The authors would also like to acknowledge the research scholars of IIT Guwahati for their participation in the perceptual test. This work is in part supported by a project entitled “NASOSPEECH: Development of Diagnostic system for Severity Assessment of the Disordered Speech” funded by the Department of Biotechnology (DBT), Govt. of India.

\balance
\bibliographystyle{IEEEtran}

\bibliography{APSIPA_CLP_ASR}

\end{document}